\documentclass[useAMS,usenatbib,twocolumn,usegraphicx]{mn2e}
\usepackage{amsmath}

\title[The fragmentation of protostellar discs]{The fragmentation of protostellar discs: the Hill criterion for spiral arms}
\author[Patrick D. Rogers and James Wadsley]{Patrick D. Rogers$^{1}$\thanks{E-mail:
rogerspd@mcmaster.ca (PDR)} and James Wadsley$^{1}$\\
$^{1}$McMaster University, 1280 Main Steet West, Hamilton, ON L8S 4M1 Canada \\
}
\begin{document}

\date{Accepted . Received ; in original form }

\pagerange{\pageref{firstpage}--\pageref{lastpage}} \pubyear{2002}

\maketitle

\label{firstpage}

%
%

\begin{abstract}
We present a new framework to explain the link between cooling and fragmentation in gravitationally unstable protostellar discs.  This framework consists of a simple model for the formation of spiral arms, as well as a criterion, based on the Hill radius, to determine if a spiral arm will fragment.  This detailed model of fragmentation is based on the results of numerical simulations of marginally stable protostellar discs, including those found in the literature, as well as our new suite of 3-D radiation hydrodynamics simulations of an irradiated, optically-thick protostellar disc surrounding an A star.  Our set of simulations probes the transition to fragmentation through a scaling of the physical opacity.  This model allows us to directly calculate the critical cooling time of \citet{Gammie2001}, with results that are consistent with those found from numerical experiment.  We demonstrate how this model can be used to predict fragmentation in irradiated protostellar discs.  These numerical simulations, as well as the model that they motivate, provide strong support for the hypothesis that gravitational instability is responsible for creating systems with giant planets on wide orbits.
\end{abstract}

%
%


%
%

\section{Introduction} \label{intro}

The fragmentation of protostellar discs through gravitational instability (GI) is a possible mechanism for the formation of gas-giant planets and brown dwarfs.  For a disc to be prone to fragmentation, there are generally thought to be two criteria that must be satisfied. 

The first criterion is that a disc must be gravitationally unstable.  This can be characterized by the Toomre $Q$ parameter \citep{ToomreQ}, which is the result of a linear stability analysis for a differentially rotating thin disc:
\begin{equation}
Q = \frac{c_s\kappa_e}{\pi G\Sigma}.
\end{equation}
In the above, $c_s$ is the sound speed of the gas, $\kappa_e$ is the epicyclic frequency ($\kappa_e = \Omega$, the rotation rate, for Keplerian rotation), $G$ is the gravitational constant, and $\Sigma$ is the surface density.  For low values of $Q \sim 1$, such as would be found in a massive, cold disc, a disc will be gravitationally unstable. 

The second criterion for fragmentation is that a disc, in addition to being gravitationally unstable, must also cool quickly.  \citet{Gammie2001}, using 2-D shearing-box simulations, examined the stability of a local patch of a protostellar disc with a simplified cooling prescription:
\begin{equation}
t_{\mbox{\tiny{cool}}}\Omega = \beta,
\end{equation}
where $t_{\mbox{\tiny{cool}}}$ is the cooling time and $\beta$ is a constant.  The cooling time here is defined through the evolution of the specific internal gas energy, $u$:
\begin{equation}
\frac{du}{dt}=-\frac{u}{t_{\mbox{\tiny{cool}}}}.
\end{equation}

By performing simulations with different values for $\beta$, the author found the cooling criterion for fragmentation to be
\begin{equation}
t_{\mbox{\tiny{cool}}}\Omega \le \beta_{\mbox{\tiny{crit}}} = 3.
\end{equation}
Subsequent work, using global simulations, has shown that the critical cooling time, $\beta_{\mbox{\tiny{crit}}}$, depends on the adiabatic index of the gas \citep*{rice2005}.  To date, the value of the critical cooling time has only been found through numerical experiment.

\citet{Gammie2001}, as well as \citet{rice2005}, outlined a physical argument for the existence of a critical cooling time.  If GI can be well-characterized by an $\alpha$-viscosity model (\citet{ShakuraSunyaev1973}; see \citet{Lodato2004, Lodato2005} for the applicability of this), then a steady-state can exist if the viscous heating by GI is balanced by the prescribed cooling.  If this balance is achieved, then the required viscosity of the disc is determined by
\begin{equation}
\alpha = \frac{1}{9/4\gamma\left(\gamma - 1\right)\beta}.
\end{equation}
If GI has a maximum $\alpha$ that it can attain, then it also has a maximum heating rate.  If the prescribed $\beta$ cools the disc faster than this maximum heating rate, then no balance between heating and cooling can be achieved, and the disc fragments.  

The $\beta$-prescription of cooling, however, is a simplified model; more generally, $\beta$ will evolve with $\alpha$ in a given disc.  In a realistic disc, heating and cooling are linked to the physical state of the disc.  In this work, we consider fragmentation in this more realistic case, whereas the $\beta$-prescription of cooling does not allow \citet{Gammie2001} to have done so.

Futhermore, the applicability of a single value of $\beta_{\mbox{\tiny{crit}}}$ (or $\alpha_{\mbox{\tiny{max}}}$) is in some doubt.  \citet{Meru2011} gave an overview of some of the inconsistencies in the literature regarding the use of a single value for the critical cooling time.  From a set of their own simulations, the authors found that a single value for the critical cooling time was not applicable; instead, they found an empirical relation in which $\beta_{\mbox{\tiny{crit}}}$ is a function of the distance from the central star, the local surface density, and the stellar mass.  However, it is not clear how well this empirical relation holds, as \citet{Meru2010b} have demonstrated that 3-D simulations investigating the critical cooling time have not yet converged numerically.  

In this work, we present a set of 3-D radiation hydrodynamic simulations of a massive, optically-thick, protostellar disc, unstable near 100 AU, around an A star.  Rather than using a $\beta$-prescription for the cooling, these simulations include realistic heating and cooling of the disc, including cooling from the disc photosphere and irradiation from the central star.  We do, however, vary the cooling rate in this set of simulations by scaling the dust opacity table by a constant factor.  By reducing the opacity (which reduces the cooling time for an optically thick disc) over this set, we observe a transition from discs that are stable against fragmentation to discs that do fragment; this is consistent with the cooling criterion work of \citet{Gammie2001}.  

We have used results from \citet*{Cossins2009}, and from this set of simulations to develop a simple, yet detailed, physical model for the fragmentation of a gravitationally unstable protostellar disc.  In this model, spiral arms develop in an unstable disc on a characteristic scale related to the disc scale height \citep{Cossins2009}.  The heating rate of the disc from GI is proportional to the square of the amplitude of the surface density variations in the disc \citep{Cossins2009}; as spiral arms become more condensed, the heating rate is increased.  The cooling rate of the disc from photospheric cooling is inversely proportional to the square of the surface density; as spiral arms become more condensed, the cooling rate in the arms decreases.  There is therefore a natural scale for the thickness of a spiral arm in a gravitationally unstable disc.  This scale is set by a balance between heating from spiral waves and radiative cooling.  It is worth noting that for faster cooling rates (shorter cooling times), this thickness will be decreased.

A second scale of interest in this model is the Hill radius, which, for an object of mass $M$, is
\begin{equation}\label{eqHill}
H_{\mbox{\tiny{Hill}}} = \left[\frac{GM}{3\Omega^2}\right]^{1/3}.
\end{equation}   
The importance of the Hill radius can be understood within the context of planet formation in a disc of planetesimals around a star.  If a protoplanet embryo has formed in this disc, then it of interest to determine the radius over which it may further accrete planetesimals.  The Hill radius sets this embryo's sphere of influence: material within the Hill radius is bound to the embryo and will be accreted.  In essence, material within the Hill radius of an object is dominated by that object's gravity, while material outside of the Hill radius is dominated by the central star's gravity, which is equivalent to the role of the local shear in the Toomre criterion.  For the purpose of this discussion, we define the Hill thickness as twice the Hill radius. 

In this framework, we can extend the cooling criterion of \citet{Gammie2001} with the Hill criterion for spiral arms.  \textit{In a gravitationally unstable disc, the natural thickness of the spiral arms is set by a balance between heating and cooling.  Fragmentation occurs in this disc if there is a section of arm whose natural thickness is smaller than that section's Hill thickness.}  Essentially, if a section of a spiral arm lies within its own Hill thickness, then shear will be unable to prevent the collapse of the arm, and fragmentation can take place. 

In a gaseous disc, pressure can prevent fragmentation from taking place.  By considering the Hill radius, we have not addressed the role that pressure plays in determining fragmentation and how it may modify the critical thickness of spiral arms necessary for fragmentation to take place.  The correct determination of this scale requires the solution to a stability analysis of a spiral arm in a differentially rotating system.  Since we do not have such a solution, we have chosen to consider the Hill thickness.  The analysis of our simulations; however, does indicate that the Hill thickness is the correct scale to examine.  The Hill criterion for fragmentation is consequently an empirical criterion.

Our model is consistent with the cooling criterion: as the cooling time decreases, spiral arms become thinner and more over-dense, becoming more likely to reside within their own Hill thickness, and consequently more likely to fragment.  With the Hill criterion, however, we have developed a more detailed, and more complete, physical picture of fragmentation.  This picture can be applied to the general case of a disc with physical heating and cooling, or the more specific case of a disc with $\beta$-prescription cooling.  In fact, it offers a means to calculate what the critical cooling time is for a given region of a given disc.  

The structure of the paper is as follows.  In \S\ref{secSimulations}, we overview our numerical methods as well as our set of simulations of gravitationally unstable, irradiated protostellar discs. In section \S\ref{secFragmentationModel}, we give a detailed picture of our model of protostellar disc fragmentation and the Hill criterion.  In addition, we demonstrate the model's consistency with the simulations of \S\ref{secSimulations}.  In \S\ref{secModelPredictions}, we show that the Hill criterion is quantitatively consistent with the cooling criterion and discuss the predictive qualities of the model.  Finally, in \S\ref{secConclusions} we give our conclusions.  

%
%

\section{Numerical simulations of gravitationally unstable irradiated discs} \label{secSimulations}

\subsection{Numerical methods}

Our simulations were performed with the TreeSPH code Gasoline \citep*{gasoline}, with the addition of radiative transfer in the flux-limited diffusion approximation  [FLD; \citet{rogers2011}].  As described by the authors, FLD is able to model the transfer of energy only in regions in which SPH particles  reside.  Because of limited resolution, any SPH representation of a protostellar disc naturally has two edges, representing the upper and lower atmospheres.  Radiative cooling from the disc atmospheres is modelled by means of a photosphere boundary condition: the SPH particles on the ``edge'' of the disc (the edge-particles) are found, robust surface areas (the area of the photosphere for which an edge-particle is responsible) are calculated using a 2-D SPH estimate, and a plane-parallel cooling term is added to the radiative energy equation for the edge-particles.   The radiative hydrodynamics has been tested on a number of standard problems, including the relaxation test of \citet{Boley2007}, which is particularly suited to protostellar disc simulations.

The conditions in the outer regions of discs (roughly 100 AU and beyond) are expected to be favourable to gravitational fragmentation, since the cooling criterion is likely satisfied there \citep{Rafikov2007}.  As pointed out by \citet{Kratter2011}, the heating of the outer regions is expected to be dominated by the irradiation  of the disc's surface by the central star, rather than by viscous heating.  Since our simulations focus on fragmentation at these large radii, it is therefore fundamentally important to account for this heating via irradiation.  

The photosphere boundary condition of \citet{rogers2011} offers a straightforward means by which this can be done.  In addition to the cooling term in the specific radiation energy equation for each edge-particle, we have added a heating term of 
\begin{equation}\label{eqIrradHeating}
\left[\frac{D\xi_i}{Dt}\right]_{\mbox{\tiny{irrad}}} = \frac{A_i}{m_i}\sigma\left(T_{\mbox{\tiny{irrad}}}\right)^4,
\end{equation}
where $\frac{D}{Dt} = \frac{\partial}{\partial t} + \bf{v}\cdot \nabla$ is the co-moving derivative, $\xi$ is the specific radiation energy, $A_i$ is the surface area of the edge-particle, $m_i$ is the particle mass, $\sigma$ is the Stefan-Boltzmann constant, and $T_{\mbox{\tiny{irrad}}}$ is the temperature of the stellar irradiation.

\citet*{Kratter2010} used the passive flared disc model of \citet{ChiangAndGoldreich1997}, along with a stellar evolution model, to determine the equilibrium temperature distribution for a disc surrounding a 1.35 M$_{\odot}$ A star, which they found to be:
\begin{equation}\label{eqTirrad}
T = 40 \mbox{ K} \left(\frac{R}{70 \mbox{ AU}}\right)^{-3/7}.
\end{equation}  
Since we are not able to treat the super-heated dust layer of optical depth $\tau < 1$ in our simulations, it is appropriate to use this equilibrium temperature distribution as the $T_{\mbox{\tiny{irrad}}}$ in the irradiation heating term, equation (\ref{eqIrradHeating}).  In addition, we implement a floor  of $T_{\mbox{\tiny{irrad}}}=20$ K to take into account the background radiation field.

\subsection{Initial conditions and input parameters}

%
%

\begin{figure}
\begin{center}
\includegraphics[width=3.8in]{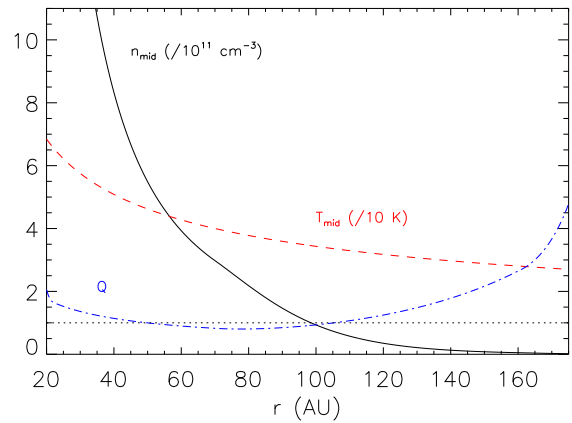} 
\end{center}
\caption{The physical quantities of the initial disc profile.  The midplane number density (in units of $10^{11}$ cm$^{-3}$) is given by the solid, black line; the midplane temperature (in units of 10 K) is given by the red, dashed line; and the $Q$ is given by the blue, dot-dashed line.  The horizontal, dotted line is a reference for $Q = 1$.}
\label{figDiscIC}
\end{figure}

The initial, axisymmetric model of $5 \times 10^{5}$ SPH particles was created in a manner similar to that of \citet{Shen2010}-- see Figure \ref{figDiscIC} for the disc properties.  The surface density profile has the form $\Sigma \propto r^{-1}$ in the region of 20-70 AU  (there is a smooth increase of $\Sigma$ from 10-20 AU).  There is a smooth functional form of $\Sigma \propto r^{-1}\exp\left\{-4\log(R)\left[0.5\log(R)-\log(R_m)\right]/\log(R_o/R_m)\right\}$, with $R_m$ = 70 AU and $R_o$ = 160 AU, from 70-160 AU, after which there is a steep drop off of $\Sigma \propto r^{-15}$.  There is roughly 0.61 M$_{\odot}$ within 200 AU.

This particular surface density distribution is motivated by the initial temperature profile, which is given by the equilibrium temperature in equation (\ref{eqTirrad}).  The combination of temperature and surface density results in a broad region of the disc having an initial $Q$ of roughly unity.  

  Using the initial surface density and temperature profiles, the vertical structure satisfying hydrostatic equilibrium was calculated iteratively taking into account both the gravity from the central star and the self-gravity of the disc.  We have chosen such a stable initial condition, with $Q \sim 1$, to ensure that the transition towards the spiral structure caused by GI is smooth and does not suffer from transients that are the result of an unstable setup.

The central star is modelled as a 1.35 M$_{\odot}$ sink particle with a radius of 10 AU.  We use a mean molecular weight of 2.3 and realistic Rosseland-mean opacities \citep*{Dalessio1997}. Although the code is capable of using a consistent treatment of the internal energy of molecular hydrogen that takes into account translational, rotational, and vibrational modes \citep{Boley2007H2}, to simplify the analysis of our simulations, we use a fixed adiabatic index of $\gamma = 7/5$.  The scale height is resolved by at least three smoothing lengths outside of 25 AU, and the Jeans length is resolved until shortly after fragmentation takes place.  We define one ORP to be the Keplerian period at our fiducial radius of 100 AU, roughly 863 years.

\subsection{Simulations}

We present a set of five simulations, each of which uses the initial conditions described above, the only difference being the opacity used.  Simulation (A, B, C, D, E) has an opacity table that is scaled by a constant value of (1/10, 1/3, 1, 3, 10).  Thus, Simulation C has an estimated physical opacity for solar metallicity, while Simulations A and B have reduced opacities, and Simulations D and E have increased opacities.  Physical changes in opacity could be the result of grain growth \citep*{Birnstiel2010}, grain evolution via the passage of spiral arms \citep*{Podolak2011}, or formation in an environment with a non-solar metallicity.   Our goal, however, is not to reproduce different physical environments, but rather to explore the necessary conditions for gravitational fragmentation.  In this context, a simple scaling of the opacity table is 
both sufficient and desirable. 


\begin{figure}
\begin{center}
\includegraphics[width=2.3in]{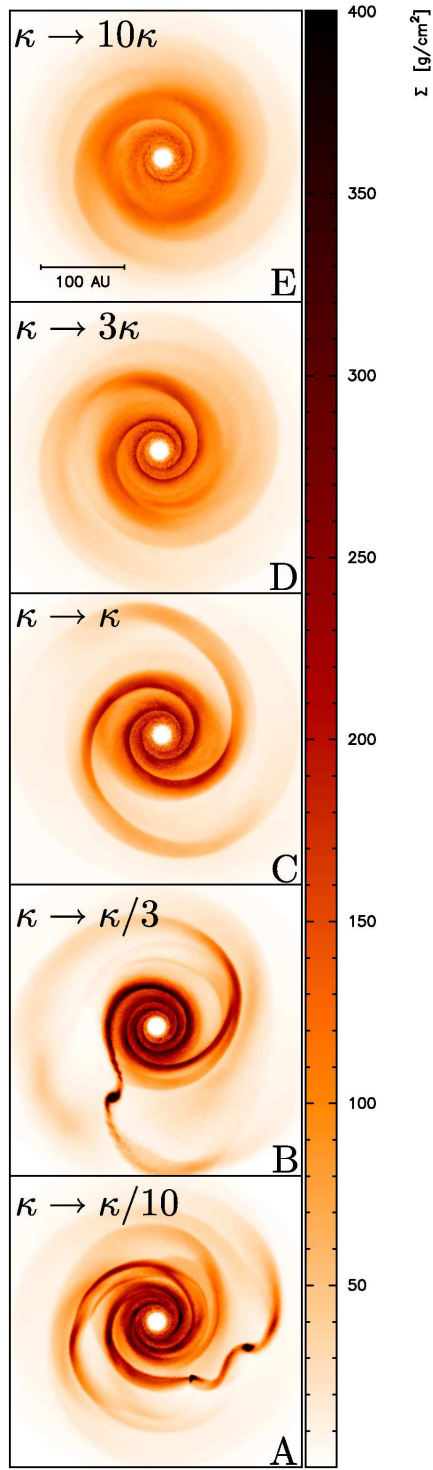} 
\end{center}
\caption{The final states of the simulated discs: surface density plots of Simulations E, D, C, B, and A are shown from top to bottom. As the opacity scaling is reduced from the physical value, fragmentation occurs. The discs are shown at respective times of 8.5, 8.5, 8.5, 3.5, and 2.9 ORPs.}
\label{figDiscEndState}
\end{figure}

All five simulations evolved in a similar fashion over the first 2.5 ORPs.  High mode-number spiral structure developed slowly from SPH Poisson noise in each of the discs over this time until settling down to a transitioning state of two or three spiral arms.  The transition from axisymmetric initial condition to spiral structure is observed to be smooth, with no strong transients.  

The final states of the simulations are shown in Figure \ref{figDiscEndState}.  Simulations C, D, and E have been evolved for roughly 8.5 ORPs without fragmentation having taken place (although strong spiral two-arm over-densities may persist), while Simulation A has fragmented with two objects forming, and Simulation B has fragmented with one object forming.  This set of simulations, therefore, demonstrates a transition from non-fragmentation to fragmentation, as a function of the opacity scaling.  

For a patch of an optically thick disk, the cooling time is approximately
\begin{equation}\label{eqtcool}
t_{\mbox{\tiny{cool}}} = \frac{1}{4}\frac{1}{\gamma-1}\frac{c_s^2\kappa}{\sigma T^4}\Sigma^2,
\end{equation}
where $\kappa$ is the opacity \citep{Rafikov2007}.  Hence, the cooling time is directly proportional to the opacity, and our set of simulations offers a means of exploring the fragmentation boundary as a function of cooling time in a manner similar to the simulations of \citet{Gammie2001}.  The difference is that our simulations use realistic radiative cooling (even though the opacities may be scaled), rather than $\beta$-prescription cooling.  

We know from the cooling criterion that reducing the cooling time (by reducing the opacity) will eventually lead to fragmentation.  Thus, that Simulations A and B fragment is consistent with this picture.  However, it is possible to use these simulations to better understand why exactly fragmentation takes place.

\begin{figure}
\begin{center}
\includegraphics[width=2.3in]{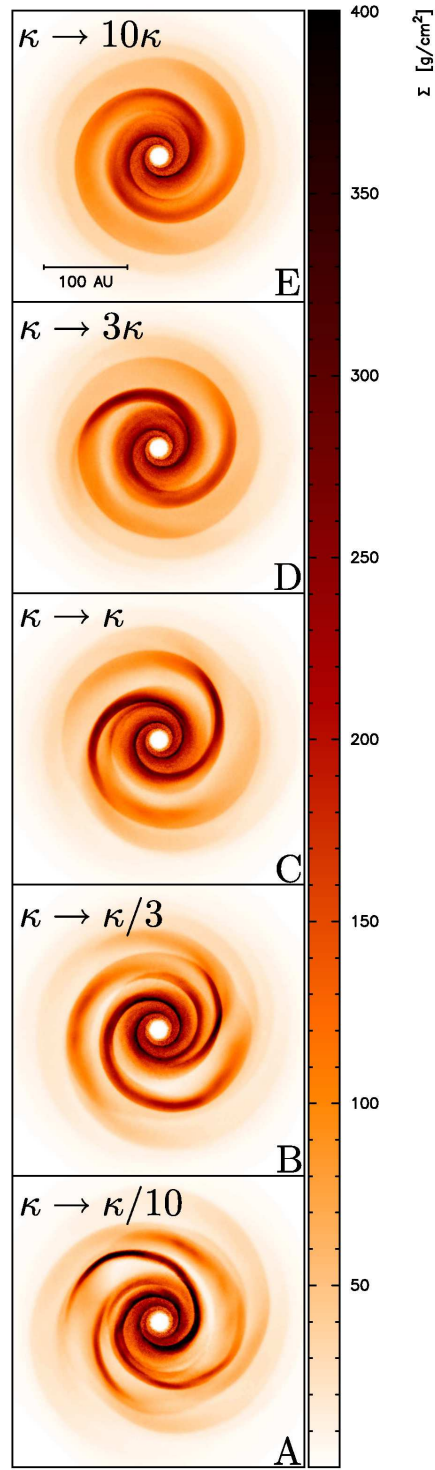} 
\end{center}
\caption{The simulated discs before fragmentation: surface density plots of Simulations E, D, C, B, and A are shown from top to bottom.  As the opacity scaling is reduced, the spiral arms become thinner and more over-dense.  The discs are shown at a time of 2.5 ORPs.}
\label{figDiscMidState}
\end{figure}

Figure \ref{figDiscMidState} shows the five simulations at roughly the same time, shortly before fragmentation took place in Simulations A and B.  The difference in the structure of the five discs at this time offers evidence of a detailed description for why Simulations A and B fragment, while Simulations C, D, and E do not.  As can be observed, as the cooling time decreases (as the opacity decreases), spiral arms in a disc become thinner and more over-dense and this makes fragmentation more likely to occur.  

%
%

\section{Fragmentation model and the Hill criterion} \label{secFragmentationModel}
We present a model of spiral arm fragmentation in gravitationally unstable discs, based on the observation from the simulations of the previous section that reduced cooling times lead to thinner arms, which are more likely to fragment.  This model can be broken into two components: the first is a model for the (roughly) steady-state spiral structure in an unstable disc; while the second is a criterion for the fragmentation of these spirals.  Many of the details of our model are empirical in nature: we have used results from the simulations of \citet{Cossins2009}, as well as our own set of simulations in determining some of the important parameters.

\subsection{Spiral structure}

\begin{figure}
\begin{center}
\includegraphics[width=3.3in]{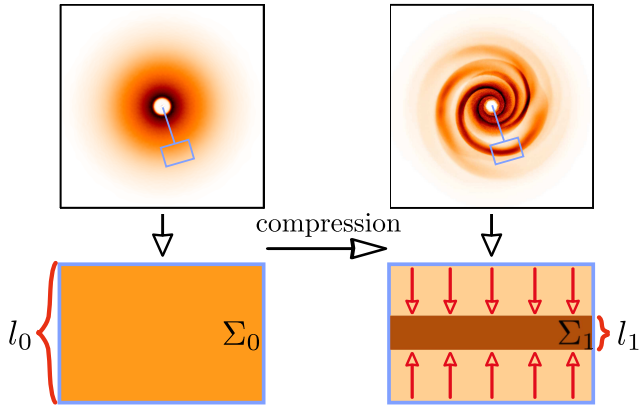} 
\end{center}
\caption{The creation of a spiral arm:  a local patch of radial extent $l_0$ (left) in an axisymmetric disc collapses radially to form a section of a tightly-wound spiral arm (right) of radial extent $l_1$.}
\label{figArmCollapse}
\end{figure}

We begin with a model for the spiral structure that results in a gravitationally unstable disc.  We consider a patch of an initially axisymmetric disc that will develop spiral structure, such as our initial condition for the simulations of the preceding section.  This patch is located at some distance, $R$, away from the central star and is of a radial extent $l_0$, with a characteristic surface density $\Sigma_0$.  GI acts on the scale of $l_0$ to collapse mass radially\footnote{The simulations of \S\ref{secSimulations} show tightly-wound spiral structure with a typical winding angle of $i \sim 10^{\circ}$; to first-order, a purely radial collapse is a fair approximation.}, resulting in the formation of a spiral arm of thickness $l_1$ and characteristic surface density $\Sigma_1$.  This process is demonstrated in Figure \ref{figArmCollapse}.  

\begin{figure}
\begin{center}
\includegraphics[width=3.3in]{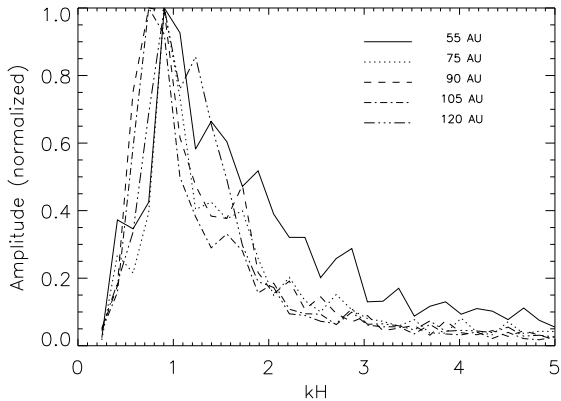} 
\end{center}
\caption{Radial Fourier analysis of Simulation B at 2.5 ORPs: the amplitude of the radial mode is given as a function of the wavenumber times the local scale height for a number of radii.  Each amplitude curve is normalized by the peak amplitude for that radius.  The Fourier analysis is carried out using the method of \citet{Cossins2009}.}
\label{figRadialFourier}
\end{figure}

What is the appropriate scale for $l_0$?  \citet{Cossins2009} performed a number of simulations of marginally stable ($Q\sim 1$) discs and found that, from a radial Fourier transform of these discs, the dominant radial wavenumber was typically
\begin{equation}
k \approx  \frac{1}{H} = \frac{\pi G \Sigma}{c_s^2}.
\end{equation}
Therefore, we expect the scale of spiral arm formation to be $l_0 = 2\pi H$.  We have tested that this is consistent with our own simulations.  Figure \ref{figRadialFourier} shows the results of a radial Fourier transform of Simulation B at a time of 2.5 ORPs.

How many spiral arms are likely to form in our disc?  Numerical studies \citep{Lodato2004, Lodato2005} have shown that as the disc-to-star mass-ratio increases, marginally unstable discs show fewer spiral arms.  Our simulations are of quite massive discs, with a disc-to-star mass-ratio of $M_d/M_\star = 0.45$.  This high disk mass is necessary for the disc to have $Q \sim 1$ near 100 AU for our realistic irradiation temperature, and results in a typical arm number of $m=2$ or $3$.

The number of arms in a disc is likely the result of swing amplification, with significant amplification of a mode, $m$, requiring the swing amplification parameter, 
\begin{equation}
X_m = \frac{\Omega^2R}{2 \pi G \Sigma m}, 
\end{equation}
to satisfy $1 < X_m < 3$ \citep{BinneyAndTremaine}.  From the above, we can see that $X_m$ roughly scales with the disc-to-star mass-ratio:
\begin{equation}
X_m \propto M_\star/M_d.
\end{equation}
Thus, for low disc-to-star mass-ratios, only high-order modes will satisfy $1 < X_m < 3$, while for high disc-to-star mass-ratios, only low-order modes will.  

What is the steady-state thickness, $l_1$, of the newly formed spiral arm?  We posit that this scale is the result of a balance between heating of the disc through the spiral waves, and radiative cooling.  Assuming that the spiral density wave deposits a fixed fraction, $\epsilon$, of its energy into the disc per dynamical time, \citet{Cossins2009} showed that the heating rate per unit mass from spiral arms can written as 
\begin{equation}\label{eqSpiralHeating}
Q^+ = \epsilon\frac{c_s^2 \mathcal M \tilde{\mathcal M} \Omega}{2}\left(\frac{\delta \Sigma}{\Sigma}\right)^2,
\end{equation}
where 
\begin{equation}
\mathcal M = \frac{m \Omega_{\mbox{\tiny{p}}}}{kc_s}   \mbox{\ and \ } \tilde{\mathcal M} = \frac{m\left(\Omega_{\mbox{\tiny{p}}}-\Omega\right)}{kc_s}
\end{equation}
are the radial phase Mach number and the Doppler-shifted radial phase Mach number, and $\Omega_{\mbox{\tiny{p}}}$ is the pattern speed.

As outlined by the authors, the pattern speed, and hence the Mach numbers, can be calculated from the dispersion relation for a finite-thickness disc:
\begin{equation}\label{eqDispersion}
m^2\left(\Omega_{\mbox{\tiny{p}}} - \Omega\right)^2 = c_s^2k^2 + \Omega^2 - \frac{2\pi G|k|}{1+|k|H},
\end{equation}
if the radial and azimuthal wavenumbers are known.  From their simulations, the authors found a relatively constant value of $\epsilon \approx 0.2$ (see their Figure 15).

In an irradiated disc, there is additional heating from the stellar irradiation, so that
\begin{equation}\label{eqTotalHeating}
Q^+ = \epsilon\frac{c_s^2 \mathcal M \tilde{\mathcal M}\Omega}{2}\left(\frac{\delta \Sigma}{\Sigma}\right)^2 + \frac{\sigma T^4_{\mbox{\tiny{irrad}}}}{\Sigma}.
\end{equation}

The spiral over-density can be calculated by assuming that some fraction, $f$, of the total mass per length within $l_0$ is compressed into the spiral arm, of thickness $l_1$:
\begin{equation}
\left(\frac{\delta \Sigma}{\Sigma}\right) = \left(\frac{l_0}{f l_1} - 1\right).
\end{equation}
 
The heating from equation (\ref{eqTotalHeating}) is balanced by radiative cooling, for which the cooling rate per unit mass, using equation (\ref{eqtcool}), is
\begin{equation}\label{eqTotalCooling}
Q^- = u/t_{\mbox{\tiny{cool}}} = \frac{4}{\gamma}\frac{\sigma T^4}{\kappa\Sigma^2},
\end{equation}
where we have used $u = c_s^2/[\gamma(\gamma-1)]$.  Setting the above cooling rate equal to the heating rate of equation (\ref{eqTotalHeating}), and using the other information in this section, as well as the initial axisymmetric properties of our disc, leaves us with an equation with only one unknown: the thickness of the spiral arm, $l_1$, in our patch of interest (assuming that we know the proper temperature for radiative cooling, $T$, in the spiral arm.  This will be elaborated upon in \S\ref{secModelPredictions}). 

\subsection{Determining fragmentation}

\begin{figure}
\begin{center}
\includegraphics[width=3.3in]{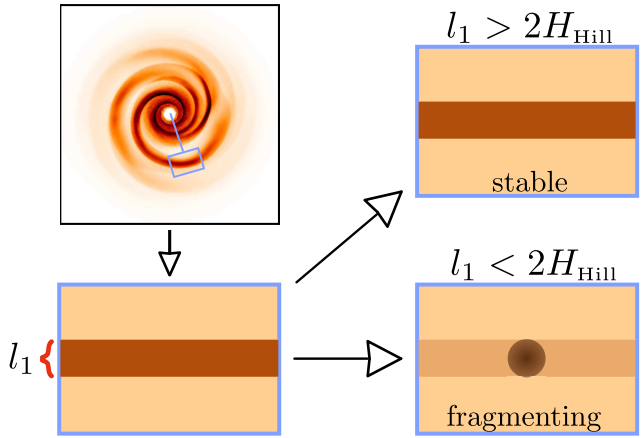} 
\end{center}
\caption{The Hill criterion for spiral arm fragmentation: if a section of spiral arm lies within its own Hill thickness, then that section of arm is free to collapse and fragmentation takes place.  If a section of spiral arm lies outside of its own Hill thickness, then shear stabilizes the arm and fragmentation does not take place.}
\label{figArmHill}
\end{figure}

Once we know the steady-state thickness of the spiral arm, $l_1$, we can use equation (\ref{eqHill}) to calculate the Hill radius for this section of the arm using:
\begin{equation}
H_{\mbox{\tiny{Hill}}} = \left[\frac{G\Sigma_1 l_1^2}{3\Omega^2}\right]^{1/3}.
\end{equation}

If the section of arm has a thickness satisfying $l_1/(2H_{\mbox{\tiny{Hill}}}) < 1$, then the section lies within its own Hill thickness.  In the absence of pressure forces, this means that the section is bound, as the tidal force from the central star (manifest as rotational shear) is less than the self-gravity of the section.  Once the section is bound, fragmentation occurs.  Conversely, if $l_1/(2H_{\mbox{\tiny{Hill}}}) > 1$, then the section of arm is not bound and fragmentation does not occur.

The Hill thickness tells one about the ability of shear to prevent the fragmentation of the arm and is therefore expected to be an important scale. However, in comparing the radial thickness of the arm to the Hill thickness we are ignoring the role of pressure, despite strong radial pressure gradients present across the arm.  It is therefore reasonable to expect that the critical thickness for arm fragmentation may be modified from the Hill thickness.  There are no strong pressure gradients along the arm (the azimuthal direction) though, so it is likely that fragmentation occurs in this direction.  In this case, the Hill thickness may in fact be the critical scale.  

Determining the correct scale for fragmentation requires a detailed calculation of the stability of a spiral arm accounting for differential rotation.  In the absence of such a calculation, we posit that the correct scale to consider is indeed the Hill thickness.  As described below, the results from our simulations are consistent with this.  The Hill criterion for fragmentation, demonstrated in Figure \ref{figArmHill}, is thus empirically based.

\subsection{Consistency with simulations}

%
%

\begin{figure}
\begin{center}$
\begin{array}{cc}
\includegraphics[width=3.5in]{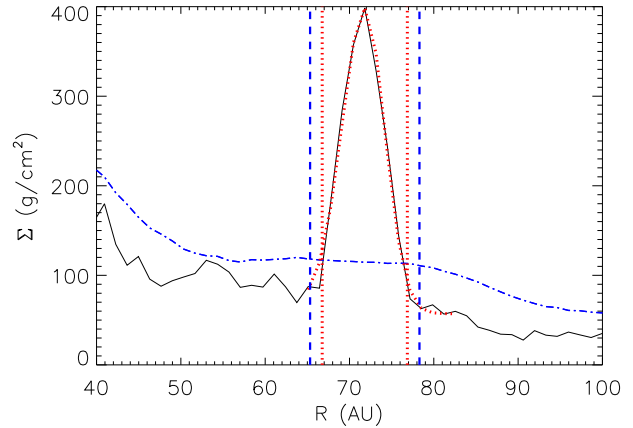} \\
\includegraphics[width=3.5in]{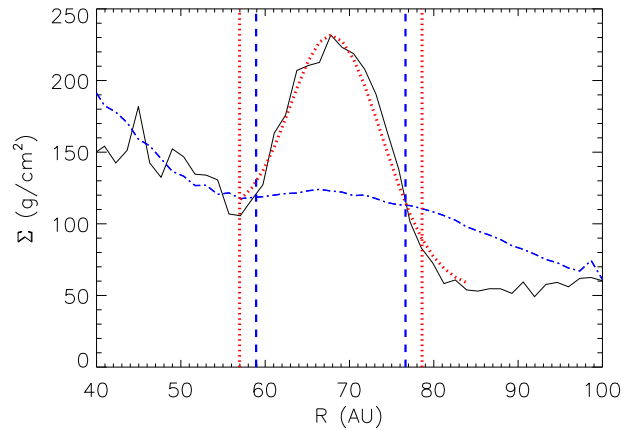} 
\end{array}$
\end{center}
\caption{Comparison of a spiral arm's thickness to its Hill thickness.  A radial slice of Simulation A (Simulation E) is shown in the top (bottom) panel.  The surface density of the radial slice is given by the black line, while the azimuthally averaged surface density is given by the blue, dot-dashed line.  As described in the text, the spiral arm (the large over-density) is fit (the red, dashed line) and the arm thickness is found (the vertical, red dashed lines).  The mass within this arm thickness is computed to determine the Hill thickness (the vertical, blue dashed lines).  Consistent with the Hill criterion, Simulation A (Simulation E) has an arm which falls inside of (outside of) its Hill thickness and fragmentation is (is not) observed shortly thereafter.}
\label{figArmThickness}
\end{figure}

An analysis of the spiral arms formed in the simulations of \S\ref{secSimulations} shows that their thicknesses, and stability, are consistent with the Hill criterion for fragmentation.  In Figure \ref{figArmThickness}, we show two examples of this analysis to illustrate this consistency.  Our analysis focuses on the surface density of a radial slice of the disc (with a typical angular width of 5$^{\circ}$).  Over this slice, a spiral arm is evident as a large over-density.  We find that arms are often asymmetric; consequently, we determine a thickness for an arm by fitting each side of the arm (with respect to the radius of highest $\Sigma$, $R_{\mbox{\tiny{peak}}}$), with a Gaussian of the form 
\begin{equation}
\Sigma_{\mbox{\tiny{arm}}} = \Sigma_0 + e^{\frac{\left(R - R_{\mbox{\tiny{peak}}}\right)^2}{2b^2}},
\end{equation}
where $\Sigma_0$ is the value of the surface density adjacent to the arm.  The thickness of the arm is taken to be $l_1  = 2(b_{\mbox{\tiny{left}}} + b_{\mbox{\tiny{right}}})$, and the mass of the section of arm is determined using a numerical evaluation of
\begin{equation}
M_{\mbox{\tiny{arm}}} = \int_{R_{\mbox{\tiny{peak}}}- 2b_{\mbox{\tiny{left}}}}^{R_{\mbox{\tiny{peak}}}+ 2b_{\mbox{\tiny{right}}}}\Theta(R)R\Sigma(R)dR,
\end{equation}  
where $\Theta(R)=l_1/R$ is the angular extent of the section of arm.

Figure \ref{figArmThickness} (top) shows this analysis for an arm in Simulation A, shortly before fragmentation.  Consistent with the Hill criterion, the arm thickness is less than the Hill thickness and therefore fragmentation is expected to occur in this arm; indeed, this arm fragmented a short time after the time-step used for this analysis.  In contrast, Figure \ref{figArmThickness} (bottom) shows this analysis for an arm in Simulation E, which never fragmented.  Consistent with the Hill criterion, the arm thickness is greater than the Hill thickness and therefore fragmentation is not expected.

We use these two examples here to illustrate the consistency of the fragmentation criterion with the simulations performed.  More generally, we have found that lower opacity discs have arms that are consistently smaller with respect to their Hill thickness in comparison with higher opacity discs.  In each of the cases where fragmentation takes place (this occurs in the reduced opacity discs), the arms that fragment are observed to fall within their own Hill thickness shortly before fragmentation.  Indeed, all arms that are observed to lie within their own Hill thickness fragment.\footnote{There is one exception: the spiral arm of Simulation C that collided with a fragment, described in the next section.}

\subsection{Comments regarding the model}\label{secComments}

The simplicity of the model described above is desirable, as it gives a straightforward, physical picture of the formation of spiral arms in a gravitationally unstable disc, as well as the physical criterion that determines whether or not those spiral arms fragment.  
However, unstable discs do exhibit a great deal of complexity; here, we discuss this complexity and comment on its implications for our model.

We have described the steady-state thickness of a single spiral arm as being the result of a balance between heating and cooling.  However, the simulations show that the spiral structure in the disc evolves with time: the number of arms in the disc is not constant, nor is the over-density of each arm.  However, an analysis of the spiral arms in these simulations shows that the radial sound crossing time of these arms is generally much less than the radiative cooling time.  As a consequence, the arms are able to adjust very quickly to perturbations, so that contractions are quasi-static.  The important point is that the arm thickness responds to the balance between heating and cooling.

%
%

\begin{figure}\vspace{40 pt}
\begin{center}
\includegraphics[width=3.0in]{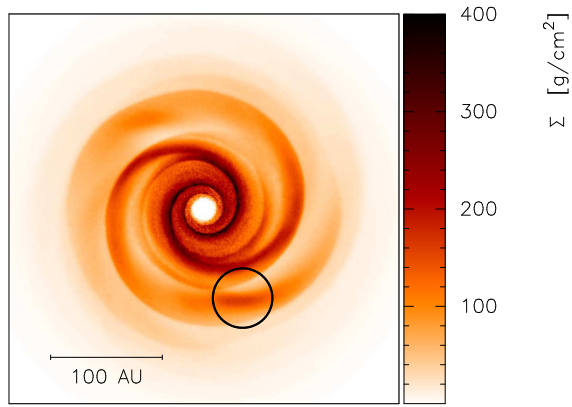} 
\end{center}
\caption{An instance of ``failed fragmentation'' in Simulation C.  In this example, a fragment, the over-density within the black circle, is observed; however, it is not long-lived.  Shortly after this time-step, the fragment collided with the spiral arm and did not survive.}
\label{figFailedFrag}
\end{figure}  

The Hill criterion describes the fragmentation of a spiral arm, but it does not necessarily determine whether or not this fragmentation leads to the formation of a long-lived object; this also depends on the cooling of the fragment and the complex environment of the disc in which the initial fragmentation takes place.   Simulation C, for example, demonstrates an instance of ``failed fragmentation,'' as shown in Figure \ref{figFailedFrag}.  One of the spiral arms appears to have fragmented; however, the resulting over-density is only short-lived:  it shortly thereafter collided with the next spiral arm, without surviving.  

The Hill criterion describes the formation of a fragment based on the inability of shear to stabilize a section of a spiral arm.  Further collapse occurs on the radiative cooling timescale of the fragment.  If this timescale is long, then the fragment may still be quite diffuse, and easily disrupted by collisions with subsequent spiral arms.  Indeed, Simulation B, which generally has shorter cooling times than Simulation C because of its opacity scaling, shows a fragment which formed, but then subsequently collided twice with spiral arms; in contrast to Simulation C, this object survived, as observed in Figure \ref{figDiscEndState}.  Fragmentation, therefore, can be well characterized by our model; however, whether or not fragments survive also depends on the complex non-linear interactions between collapsing fragments and the spiral structure in the disc.

As described, the fragment of Simulation C was disrupted through a collision with the subsequent spiral arm.  This resulted in a strong compression of the spiral arm; in fact, the compression was strong enough that the arm was observed to lie within its own Hill thickness.  Nevertheless, the arm did not fragment.  This does not conflict with the Hill criterion because in this instance, the arm was not in a near steady-state. Since the timescale for the collision was much shorter than the sound-crossing time of the arm, the arm could not adjust.  As a result,  the increased over-density of the arm lead to an increased heating rate, see equation (\ref{eqSpiralHeating}), and a reduction of the cooling time, see equation (\ref{eqtcool}).  As a consequence of the imbalance between heating and cooling, pressure forces caused the arm thickness to expand on roughly the sound-crossing time, with the result that no fragmentation took place.

%
%

\section{Consistency with the critical cooling criterion and predictive ability} \label{secModelPredictions}
The physical model developed in the preceding section can be used to examine whether or not fragmentation is likely to take place in a disc.  In this section, we demonstrate that the predictions of this model are consistent with previous numerical results of discs evolved using the $\beta$-prescription of cooling, as well as the results of our suite of simulations discussed in \S\ref{secSimulations}.  Specifically, we analyze the initial condition of our simulated disc described in \S\ref{secSimulations} and adopt general values of $l_0 = 2\pi H$, $\epsilon =0.2$, $\mathcal M \tilde{\mathcal M} = 1 $  and $f = 1$.  These values will be correct to within $\mathcal O(1)$, but will likely have variation depending on the physical properties of a disc.

\subsection{Calculating the critical cooling time}

%
%

\begin{figure}
\begin{center}
\includegraphics[width=3.5in]{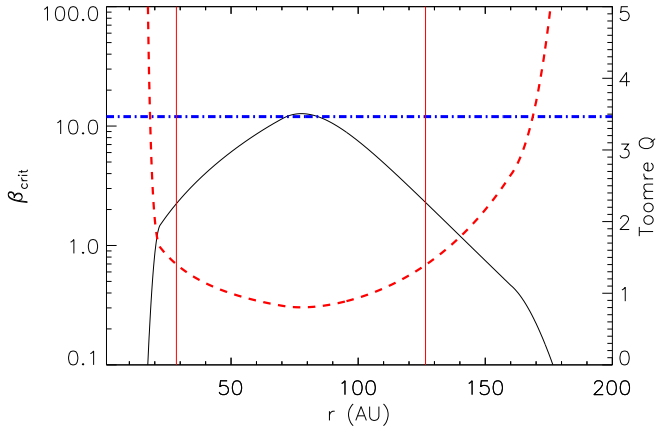}
\end{center}
\caption{The critical cooling time (black line), as calculated for our disc initial condition.  The horizontal blue, dot-dashed line represents the value as previously determined by numerical experiments, while the red, dashed line shows the $Q$ profile.  The vertical, red lines show the extent of the unstable region with $Q<1.4$.  As can be seen, the value for $\beta{\mbox{\tiny{crit}}}$ in this region is consistent with the value from numerical experiment.}
\label{figBetaCrit}
\end{figure}

The critical cooling time, $\beta_{\mbox{\tiny{crit}}}$, for a marginally stable disc with $\gamma=7/5$ has been found to be $\beta_{\mbox{\tiny{crit}}}=12$ from numerical experiments [\citet{rice2005}; with the caveat that numerical convergence has not been clearly demonstrated].  If we adopt a heating rate without irradiation (consistent with the aforementioned simulations), given by equation (\ref{eqSpiralHeating}) and balance this heating with a $\beta$-prescription cooling rate, given by 
\begin{equation}
Q^{-} = \frac{c_s^2\Omega}{\gamma (\gamma-1)\beta},
\end{equation}
then we can solve for the cooling rate that results in a certain arm thickness: 
\begin{equation}
\beta = \frac{2}{\epsilon \mathcal M \tilde{\mathcal M} \gamma\left(\gamma-1\right)}\left[1 - \frac{l_0}{f2^{3/2}}\left(\frac{2H_{\mbox{\tiny{Hill}}}}{l_1}\right)^{3/2} \left(\frac{3f\Omega^2}{G\Sigma_0 l_0}\right)^{1/2}\right]^{-2}.
\end{equation}
The critical cooling time can then be computed from the above equation by setting the arm thickness to exactly match the Hill thickness, $l_1/(2H_{\mbox{\tiny{Hill}}}) = 1$.  

The critical cooling time computed for our disc initial condition is shown in Figure \ref{figBetaCrit}.  As can be seen, we do not find a unique value for the critical cooling rate, but rather a value that depends strongly on radius (due to the variation of disc properties with radius).  Importantly, we observe that our model of disc fragmentation predicts critical cooling times in the unstable region of the disc (where $Q \sim 1$) that are consistent with the results of numerical experiments.

This consistency is a useful check on our model.  In addition, it is noteworthy that Figure \ref{figBetaCrit} demonstrates the first calculation of the critical cooling time from a physical model of fragmentation.  Previous estimates of $\beta_{\mbox{\tiny{crit}}}$ have come only from numerical experiments.  

\subsection{Predictive ability of the model for irradiated discs}

It is useful to check our model against the results of previous work using cooling in the form of a $\beta$-prescription. However, it is of particular interest to apply the model to the more realistic case of an irradiated disc with radiative cooling.  Without considering GI, an irradiated disc has a natural equilibrium state in which the heating of a particular patch of disc from stellar irradiation is balanced by the cooling of that patch from the radiative cooling of the disc photosphere.  Here, we consider deviations from this equilibrium state due to GI.  

%
%

\begin{figure}
\begin{center}
\includegraphics[width=3.5in]{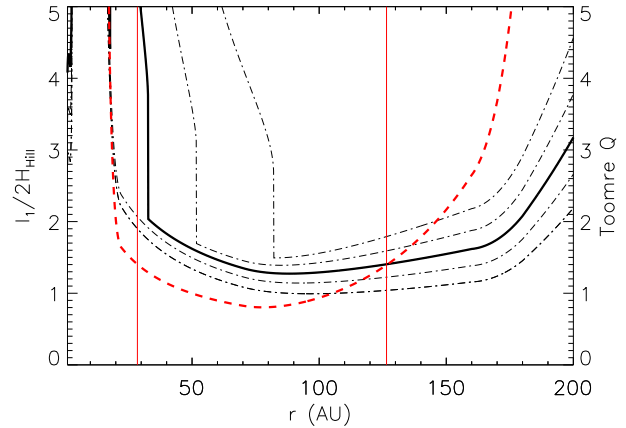} 
\end{center}
\caption{The spiral arm thickness (black curves) of an irradiated disc in units of the Hill thickness, as calculated for our disc initial condition.  The arm thickness is calculated for the range of opacities used in the simulations.  From the curve of greatest $l_1/(2H_{\mbox{\tiny{Hill}}})$ to the curve of smallest $l_1/(2H_{\mbox{\tiny{Hill}}})$ the opacity scalings are 10, 3, 1, 1/3, and 1/10 the physical opacity (the solid black curve).  The $Q$ profile is given  by the red, dashed curve, while the vertical, red lines show the extent of the unstable region with $Q<1.4$. }
\label{figl1On2Hill}
\end{figure}

Specifically, there is an additional heating of the disc from the spiral arms, given by equation (\ref{eqSpiralHeating}), which will result in an increase in temperature of $\delta T$.  We consider spiral arms in which this excess heating is balanced by a perturbative radiative cooling; that is, a cooling rate for the dissipation of this excess thermal energy. The perturbative cooling time for an irradiated disk, when the temperature in the arm is comparable to the irradiation temperature (as we find for the discs in \S\ref{secSimulations}), is \citep[e.g.][]{Kratter2010}
\begin{equation}
 t_{\mbox{\tiny{cool}}} = \frac{3}{32}\frac{\gamma}{\gamma-1}\frac{c_s^2}{\sigma T_{\mbox{\tiny{irrad}}}^4}\kappa\Sigma^2\label{eqKrattcoool}.
\end{equation}

Balancing the perturbative heating and cooling, we solve for the natural arm thickness for our disc initial condition from \S\ref{secSimulations} in units of the Hill thickness.  This is done by solving the following quartic:
\begin{equation}\label{eqQuartic}
\left\{C\right\}l_1^4 - \left\{B\right\}l_1^2 + \left\{\frac{2Bl_0}{f}\right\}l_1 - \left\{B\left(\frac{l_0}{f}\right)^2\right\}= 0,
\end{equation}
where
\begin{align}
B &= \frac{\epsilon c_s^2\mathcal M \tilde{\mathcal M}\Omega}{2} \\
\mbox{and} \nonumber \\
C &= \frac{32\sigma T^4_{\mbox{\tiny{irrad}}}}{3\kappa\gamma}\left(\frac{f}{\Sigma_0 l_0}\right)^2.
\end{align}
Once $l_1$ is calculated, we can calculate the ratio
\begin{equation}
\left(\frac{l_1}{2H_{\mbox{\tiny{Hill}}}}\right) = \frac{l_1^{2/3}}{2}\left[\frac{3f\Omega^2}{G\Sigma_0 l_0}\right]^{1/3}.
\end{equation}
The results of these calculations, for the range of opacity scalings used in our simulations, are shown in Figure \ref{figl1On2Hill}.

We expect discs to fragment for $l_1/(2H_{\mbox{\tiny{Hill}}}) \le 1$, and the trend in Figure \ref{figl1On2Hill} is consistent with this picture: as found in our simulations, the increased-opacity discs are less likely to fragment (have larger $l_1/(2H_{\mbox{\tiny{Hill}}})$) than the reduced-opacity cases.  However, even though the increased-opacity discs are not expected to fragment, the arm thickness is expected to be within a factor of two of the Hill thickness in the region of $Q\sim 1$.  This result is consistent with our simulations, as is demonstrated for the arm from Simulation E shown in Figure \ref{figArmThickness} (bottom).  This arm is stable, in that it does not lie within its Hill thickness; however, its thickness only exceeds the Hill thickness by a relatively small factor.

From our calculation, only the lowest opacity case, corresponding to Simulation A, has $l_1/(2H_{\mbox{\tiny{Hill}}}) < 1$, and would be expected to fragment; in fact, both Simulation A, and the second-lowest opacity case, Simulation B, have shown fragmentation.  This discrepancy is likely simply the result of the choice of parameters used in the calculation.  We have chosen values for a number of parameters in our model ($l_0$, $f$, $\epsilon$, and $\mathcal M \tilde{\mathcal M}$) that are expected to be correct to within $\mathcal O(1)$; however, the exact values will likely have some variation.  With an improved understanding of the growth of spiral structure, and the heating of spiral arms, the model's predictive abilities will be improved.


In the inner regions of irradiated discs, our application of a perturbative cooling is not expected to be appropriate.  In these regions, the heating is expected to be dominated by viscous heating, rather than by irradiation \citep{Kratter2011}, which may result in a disc temperature that is significantly different from the irradiation temperature.  The breakdown of our calculation can be observed as the vertical jumps in  $l_1/(2H_{\mbox{\tiny{Hill}}})$ observed in Figure \ref{figl1On2Hill}, such as the jump at 80 AU for the highest opacity disc.  Inside of this jump radius, our calculations do not give the correct result.  It is possible to calculate the proper $l_1/(2H_{\mbox{\tiny{Hill}}})$ in this region by balancing the total heating and cooling from equations (\ref{eqTotalHeating}) and (\ref{eqTotalCooling}), rather than the perturbative variants; however, this requires a knowledge of the temperature within the spiral arms.  Therefore, we leave an analysis of these regions for future work.


%
%

\section{Discussion and conclusions} \label{secConclusions}

\subsection{Implications for planet formation}
Direct-imaging observations have shown the existence of gas-giant planets at large distances from their host A star, including HR 8799b, 7 M$_{\mbox{\tiny{Jup}}}$ at a distance of  68 AU \citep{Marois2008},  and Fomalhautb, 3 M$_{\mbox{\tiny{Jup}}}$ at a distance of 119 AU \citep{Kalas2008}.  It is difficult to explain the existence of gas-giants at such distances from their host star in the core accretion scenario, since the surface densities are typically too low to form the necessary rocky cores within the lifetime of the gas disc \citep{Dodson-Robinson2009, Rafikov2011}.  However, more investigation is warranted in order to determine if such planets can be explained in the core-accretion scenario.  
ø
In comparison, fragmentation via GI has been shown to be a viable formation mechanism at large distances from the host star from both theoretical arguments \citep{Rafikov2007, NeroB2009, Kratter2010}, as well as from numerical simulations with radiative transfer \citep{Boley2009, Stamatellos2011, Boss2011}.  

The particular set of 3D radiation hydrodynamic simulations presented here were designed to investigate fragmentation at large radii ($\sim$ 100 AU) around A stars.  At these distances, heating from stellar irradiation is expected to be the dominant heating source; we have included irradiation using the $T_{\mbox{\tiny{irrad}}}$ expected for a 1.35 M$_{\odot}$ A star \citep{Kratter2010}.  

The results of our simulations show that GI can produce gravitationally bound objects at large distances from the star, given opacities on the low side of the expected range.  Such opacities could be the result of grain-growth \citep{Birnstiel2010}, grain evolution via the passage of spiral arms \citep{Podolak2011}, or formation in an environment with a non-solar metallicity \citep[HR 8799 is roughly 1/3 solar metallicity,][]{Marois2008}.  Our simulations do not take these physical mechanisms into account, but rather use a simple scaling of the opacity table.

Although our simulations do not have the resolution to follow the evolution of bound objects as their central densities run away, it is interesting to consider the objects at the end-state of our simulations, shown in Figure \ref{figDiscEndState}.  At the end of Simulation A (with an opacity scaled by 1/10), there are two brown dwarfs of masses 21 M$_{\mbox{\tiny{Jup}}}$ and 15 M$_{\mbox{\tiny{Jup}}}$, at respective distances of 62 AU and 95 AU; while at the end of Simulation B (with an opacity scaled by 1/3), there is one brown dwarf of mass 40 M$_{\mbox{\tiny{Jup}}}$ at a distance of 95 AU.  Neither the masses nor the distances of these objects represent their final state: all of the objects are accreting mass and migrating inwards at the end of the simulation.  

We conclude that GI in unstable discs can produce brown dwarfs at large distances from A stars.  We have, of course, only shown fragmentation for a single surface density and temperature profile.  It is of interest to investigate a greater region of the parameter space with numerical simulations in order to explore the possibility of low-mass companions such as those observed by \citet{Marois2008} and \citet{Kalas2008}.

\subsection{Physical model of fragmentation}
We have presented a new framework to explain the link between cooling and fragmentation in protostellar discs.  This framework consists of two components.  The first is a simple model for the formation of spiral arms, in which the thickness of a spiral arm is set by a balance between heating (through gravitational instability and irradiation) and radiative cooling.  The second is a criterion for fragmentation: spiral arms that have a natural thickness smaller than their Hill thickness fragment, resulting in objects that may survive to become gas-giant planets or brown dwarfs.

This model of fragmentation is based on results from \citet{Cossins2009} as well as our suite of 3-D radiation hydrodynamics simulations of gravitational instability in an irradiated, optically-thick protostellar disc surrounding an A star.  By reducing the opacity scaling, and consequently the cooling time, over the set of simulations, we have produced a suite that demonstrates the transition from non-fragmentation to fragmentation.  From an analysis of these simulations, we have found that the critical scale for determining fragmentation is roughly the Hill thickness: those spiral arms that are found to fragment lie within twice their Hill radius, while those spiral arms that do not fragment extend beyond their Hill thickness.  In the future, it would be of interest to have a robust calculation of the critical scale for fragmentation from a stability analysis of a spiral arm in a differentially rotating system.

In comparison to the critical cooling time picture, our model of fragmentation is a more detailed, and more general, physical picture of fragmentation that is applicable to discs with realistic heating and cooling, in addition to discs with $\beta$-prescription cooling.  Indeed, by coupling the Hill criterion to our simple model of spiral arm formation, heating, and cooling using a $\beta$-prescription, we have been able, for the first time, to calculate $\beta_{\mbox{\tiny{crit}}}$.  We find that there is not a single value for $\beta_{\mbox{\tiny{crit}}}$, but that it depends on the local properties of the disc; in addition, our calculation is consistent with the value determined by numerical experiment.  

We have also demonstrated how this model can be used to predict fragmentation in irradiated discs with radiative cooling.  Applying the model to the initial condition of our simulated disc, for the various opacity scalings used, yields predictions that are consistent with the results of our simulations.  An improvement in the predictive abilities of the model depends on a better understanding of several parameters that describe the formation and heating of the spiral arms.  

This model has been developed in the context of protostellar discs; however, it may also be of use in the context of star formation in a disc near the Galactic centre \citep{Levin2003}, as well as star-cluster formation in optically-thick starburst galaxies, such as Arp 220.  It is beyond the scope of this work to consider these systems, and we leave these considerations to future work.

In this work, we have considered isolated discs; that is, the effects of accretion from the surrounding envelope were ignored.  However, accretion is expected to play an important role in gravitationally unstable discs [\cite{Boley2009, Kratter2010a}], since it is accretion that will push the mass of the disc towards being sufficient for instability to set in, keep it unstable despite mass-transport, and contribute to heating.  In future work, we intend to investigate the effects of accretion on the stability of protostellar disks, in the context of our model of fragmentation.

%
%

\section{Acknowledgements}
We would like to thank Ralph Pudritz for useful discussions.  Surface density plots were created using SPLASH \citep{SPLASH}.  This work was supported by the National Sciences and Engineering Research Council of Canada (NSERC).  Computations were carried out using the facilities of the Shared Hierarchical Research Computing Network (SHARCNET), which is located in Ontario, Canada.

%
%

\bibliography{references}

\label{lastpage}

\end{document}